\documentclass[12pt,preprint]{aastex}

\usepackage{amsmath}

\shorttitle{Validating Time-Distance Far-Side Imaging}
\shortauthors{Hartlep et al.}

\begin{document}

\title{Validating Time-Distance Far-side Imaging of \\ Solar Active Regions through Numerical Simulations}
\author{Thomas Hartlep\altaffilmark{1}, Junwei Zhao\altaffilmark{2}, Nagi N. Mansour\altaffilmark{1}, and Alexander G. Kosovichev\altaffilmark{2}}
\altaffiltext{1}{NASA Ames Research Center, M/S 230-2, Moffett Field, CA 94035-1000}
\altaffiltext{2}{W. W. Hansen Experimental Physics Laboratory, Stanford University, Stanford, CA 94305-4085}


\begin{abstract}

Far-side images of solar active regions have become one of the
routine products of helioseismic observations, and are of importance
for space weather forecasting by allowing the detection of sunspot
regions before they become visible on the Earth side of the Sun. An
accurate assessment of the quality of the far-side maps is
difficult, because there are no direct observations of the solar
far side to verify the detections. In this paper we assess far-side
imaging based on the time-distance helioseismology method, by using
numerical simulations of solar oscillations in a spherical solar
model. Localized variations in the speed of sound in the surface and
subsurface layers are used to model the perturbations associated
with sunspots and active regions. We examine how the accuracy of the
resulting far-side maps of acoustic travel times depends on the size
and location of active regions. We investigate potential artifacts
in the far-side imaging procedure, such as those caused by the
presence of active regions on the solar near side, and suggest how
these artifacts can be identified in the real Sun far-side images
obtained from SOHO/MDI and GONG data.

\end{abstract}



\section{INTRODUCTION}

Imaging of active regions on the far side of the Sun, the side
facing away from Earth, is a valuable tool for space weather
forecasting, as well as for studying the evolution of active
regions. It allows monitoring of active regions before they rotate
into the near side, and after they rotate back into the far side.
Far-side images are produced daily by the acoustic holography
technique \citep{lin00a} using observations from both the Global
Network Oscillation Group (GONG), and the Michelson Doppler Imager
(MDI) on board the Solar and Heliospheric Observatory (SOHO).
\citet{lin00a} and \citet{duv01} have pioneered the solar far-side
imaging work, mapping the central region of the far-side Sun by
analyzing acoustic signals for double-skip ray paths on both sides
of the far-side region, by use of helioseismic holography \citep[for
a review, see][]{lin00b} and time-distance helioseismology \citep{duv93}
techniques, respectively. \citet{bra01} further
developed their technique to map the near-limb and polar areas of
the solar far side by combining single- and triple-skip acoustic
signals. More recently, \citet{zha07} has developed a five-skip
time-distance imaging scheme that measures travel times of a combination of double-
and triple-skip acoustic wave signals. Combined with the
traditionally used four-skip far-side imaging schemes, the new
technique greatly reduces the noise level in far-side images,
as well as helps to remove some of the spurious features visible
in the four-skip images.

In general terms, far-side imaging by time-distance helioseismology detects changes in the travel time for acoustic waves traveling through an active region compared to those traveling only in the quiet Sun, while helioseismic holography detects phase shifts in acoustic wave signals.
The exact mechanisms of how the presence of an active region causes the observed variations are not fully
understood, although, it is generally believed that a change of the magnetoacoustic wave speed inside active regions \citep{kosovichev1997,kos00,zha06} plays an important role.
Also, it has been argued that strong surface magnetic fields associated with active regions \citep{fan95,lin05} may affect inferences obtained by the acoustic holography technique; however, \citet{zha06} have shown that these effects are not a major factor in the determination of the interior structure of sunspots by time-distance helioseismology.

Far-side imaging has been successful for predicting the appearance
of large active regions and complexes of activity.  However, it is
unclear how robust and accurate the far-side imaging techniques are,
and how much we should believe in the far-side images that are being
produced daily. Past efforts have tried to evaluate the accuracy of
far-side images by comparing them with the directly observed
Earth-side images just after active regions have rotated into view
from the far side, or before they have rotated out of view into the
far side \citep{gonzales2007}. However, active regions may
develop quite fast, emerging or disappearing on a time scale of
days or even less. Therefore, such analyses are not sufficient.
On the other hand, numerical modeling of solar oscillations can provide
artificial data that can enable evaluating and improving these
methods. In a global solar model, we can place near-surface
perturbations mimicking active regions on the far side of the
modeled Sun, and apply helioseismic imaging techniques to the
simulated wavefield. The resulting far-side images can be
compared directly with the precisely known properties of the
perturbations, allowing for a more accurate evaluation of the
capabilities and limitations of the far-side imaging techniques.

In this paper, we present results on testing the recently
improved time-distance helioseismology far-side imaging technique
\citep{zha07} by using 3D numerical simulations of the oscillations
in the global Sun. We assess the sensitivity of the imaging
technique by varying the size and location of a sound speed
perturbation mimicking a single active region. In other simulations,
we place two active regions at the solar surface in order to examine
whether the acoustic waves traveling through the active regions may
interfere with each other and affect the imaging of the other
region. Finally, we identify one scenario in which artifacts (``ghost images'')
caused by an active region on the near side appear in the
far-side maps. A brief description of the simulation technique is
given in \S~2, followed by a description of the far-side imaging
procedure in \S~3. The main results are presented in \S~4, and a
discussion and concluding remarks are given in \S~5.


\section{NUMERICAL SIMULATION}


\subsection{Simulation Code}

In the following, we briefly describe the numerical simulation code used in this study.
For more details, the reader is referred to \citet{har05}, and in particular, to a detailed description of the code,
which will be published soon
(Hartlep \& Mansour, in preparation).

Simulating the 3D wavefield in the full solar interior is not an easy task, and many simplifications have to be made to make such
simulations feasible on currently available supercomputer systems.
For the present case, we model solar acoustic oscillations in a spherical domain by using linearized Euler equations and consider a
static background in which only localized variations of the sound speed are taken into account.
The oscillations are assumed to be adiabatic, and are driven by randomly forcing density perturbations near the surface.
For the unperturbed background model of the Sun, we use the standard solar model S of~\citet{chr96} matched to a model for the chromosphere~\citep{ver81}.
Localized sound speed perturbations of various sizes are added in the surface and subsurface layers to mimic the perturbations of the wave speed associated with sunspots and active regions.
Non-reflecting boundary conditions are applied at the upper boundary by means of an absorbing buffer layer with a damping coefficient that is zero in the interior and increases smoothly into the buffer layer.

The linearized Euler equations describing wave propagation in the Sun are written in the form:
\begin{eqnarray}
  \label{Eq:E1}
  \partial_t \rho^\prime & = & - \Phi^\prime + S - \chi \rho^\prime,  \\
  \label{Eq:E2}
  \partial_t \Phi^\prime & = & - \Delta c_0^2 \rho^\prime + \nabla \cdot \rho^\prime \mathbf{g_0} - \chi \Phi^\prime,
\end{eqnarray}
where $\rho^\prime$ and $\Phi^\prime$ are the density perturbations and the divergence of the momentum perturbations associated with the
waves, respectively.
$S$ is a random function mimicking acoustic sources, $c_0$ is the background sound speed, $\mathbf{g_0}$ is the acceleration due to gravity, and $\chi$ is the damping coefficient of the absorbing buffer layer.
Perturbations of the gravitational potential have been neglected, and the adiabatic approximation has been used.
In order to make the linearized equations convectively stable, we have neglected the entropy gradient of the background model.
The calculations show that this assumption does not significantly change the propagation properties of acoustic waves including their frequencies, except for the acoustic cut-off frequency, which is slightly reduced.
This is quite acceptable for our purpose, because the part of the spectrum that is actually used in the far-side imaging technique lies well below this cut-off frequency.
For comparison, other authors have modified the solar model including its sound speed profile~\citep[e.g.][]{hanasoge2006,par07}.
In those cases, the oscillation mode frequencies may differ significantly from the real Sun frequencies.

Starting from Eqs.~(\ref{Eq:E1}) and~(\ref{Eq:E2}), we absorb the damping terms $\chi \rho^\prime$ and $\chi \Phi^\prime$ into the other terms by use of an integrating factor, and apply a Galerkin scheme for the numerical discretization.
Spherical harmonic functions are used for the angular dependencies, and 4th order B-splines~\citep{lou97,kra99} for the radial direction.
2/3-dealiasing is used in the computations of the $c_0^2\rho^\prime$-term in angular space, while all other operations are performed in spherical harmonic coefficient space.
The radial resolution of the B-spline method is varied proportionally to the local speed of sound, i.e. the generating knot points are closely space near the surface (where the sound speed is small), and are coarsely spaced in the deep interior (where the sound speed is large).
The simulations presented in this paper employ the spherical harmonics of angular degree $l$ from 0 to 170, and 300 B-splines in the radial direction.
A staggered Yee scheme~\citep{yee66} is used for time integration, with a time step of 2 seconds.

\begin{figure*}
  \epsscale{0.9}
  \plotone{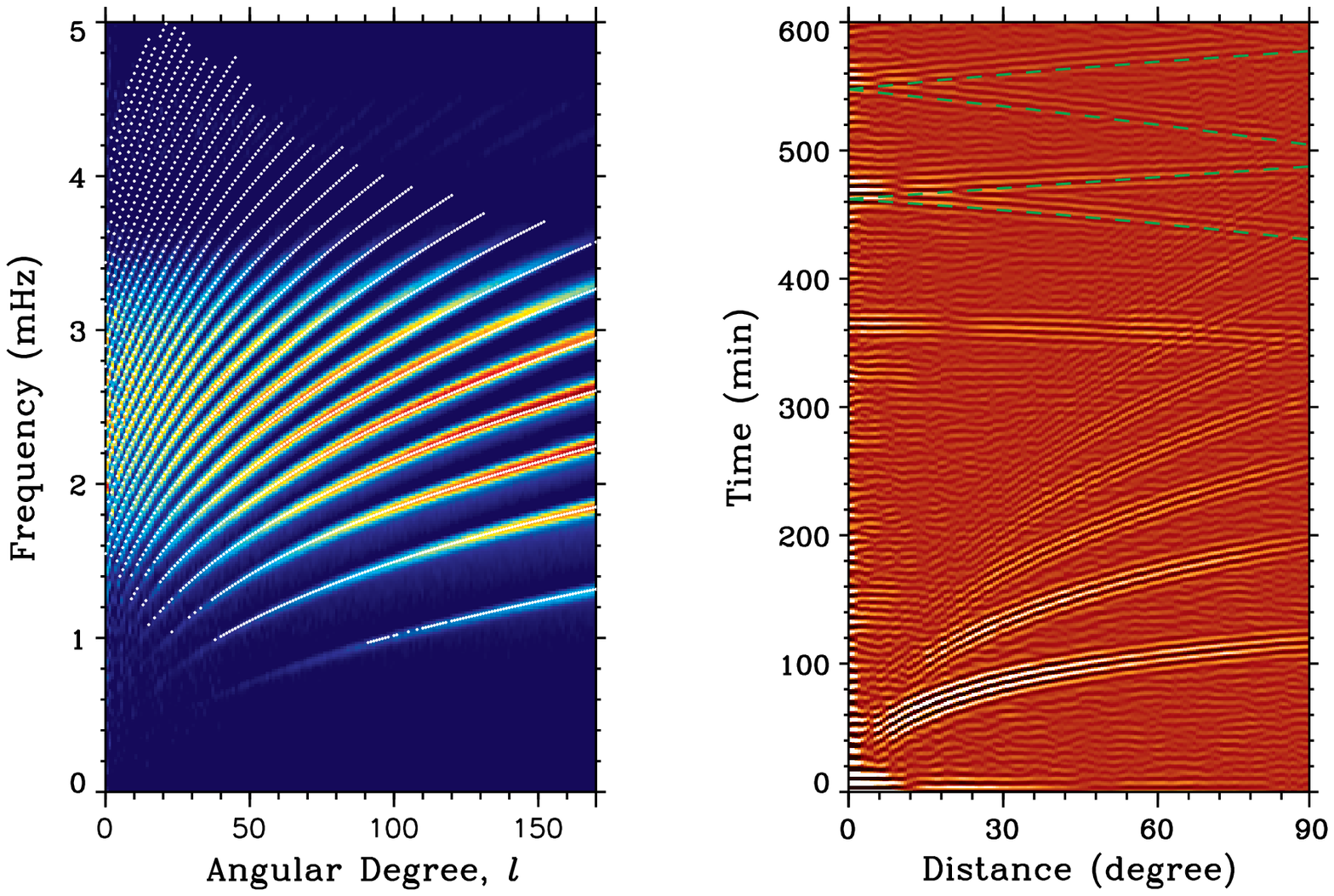}
  \caption{The oscillation power spectrum ({\it left\/}) and the time-distance diagram
  ({\it right\/}) of a simulated data set.
The white dots in the left panel show for comparison the observed
frequencies obtained from 144 days of MDI medium-$l$ data using the
averaged-spectrum method~\citep{rho97}. The green dashed lines in
the right panel indicate the ray-theory predictions for the
four-skip and five-skip signals of the acoustic wave packets, which
travel from the Earth side to the far side and back to the Earth
side}
  \label{fg1}
\end{figure*}

The oscillation power spectrum as a function of spherical harmonic degree $l$ computed for one of the performed simulations is shown in Figure~\ref{fg1}.
It is found that the frequencies of the ridges correspond well with the frequencies from solar observations.
As noted before, the model has a lower cut-off frequency, but this does not pose a problem for our purposes.
Also, Figure~\ref{fg1} shows a time-distance diagram (i.e., the mean cross-covariance function) calculated for the same simulation data.
Even though no filtering has been done for computing the time-distance correlations, both the four-skip and five-skip acoustic signals needed for the far-side imaging technique are clearly visible.
In fact, these correlations are stronger than in observational data, where it is essential to filter out other unwanted wave components \citep[compare, e.g.][]{zha07}.
The acoustic travel times are fairly close to those found in solar observations.
Even for the long travel times of four- and five-skip signals, the discrepancy between the simulations and the observations is only about 1.2~minutes, or 0.2~percent.


\subsection{Active Region Model}

\begin{figure}
  \epsscale{0.8}
  \plotone{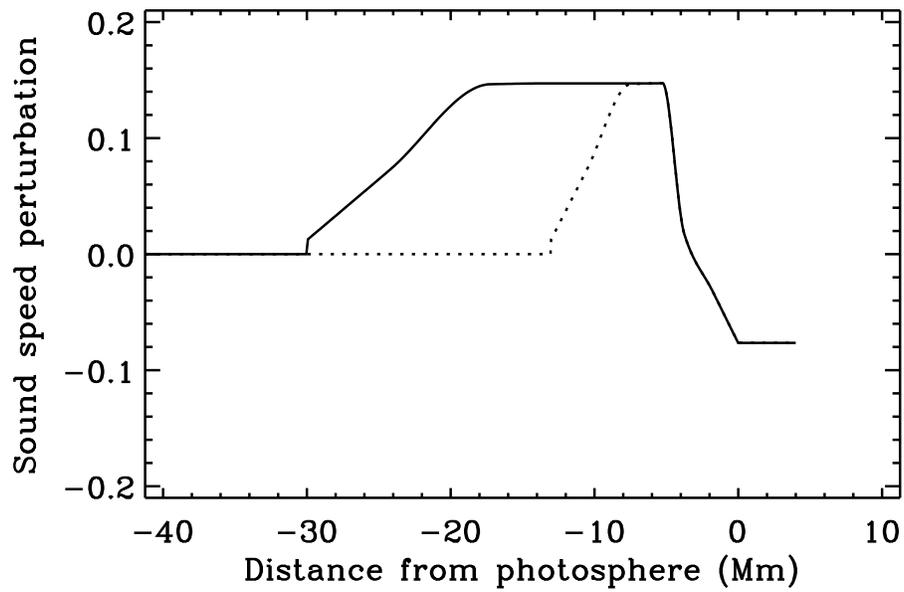}
  \caption{The radial profile of the sound speed perturbation in the center
  of the model active region ({\it solid curve\/}), with positive distances
  denoting locations above the photosphere.
The profile was derived by extending the perturbation profile of a
sunspot in NOAA active region 8243 on June 18, 1998 ({\it dotted
curve\/}) inferred from time-distance measurements \citep{kos00}.
The profile is extended to account for deeper perturbations of large active regions (see text for details).}
  \label{fg2}
\end{figure}

Solar active regions are complex structures and are believed to differ from the quiet Sun in their temperature, density, and sound speed distributions, and include complicated flow and magnetic field configurations.
The acoustic wave speed variations inside active regions due to temperature changes and magnetic fields has, for obvious reasons, a very direct effect on the travel times.
For this investigation, we model active regions by local sound speed perturbations, which include the combined temperature and magnetic effects \citep{kos00}, but leave it to a later investigation to include plasma flows.
Since the main goal of the current far-side imaging efforts is to detect the locations of active regions and estimate their size, this is quite sufficient.
We model a solar active region by a circular region in which the sound speed $c$ differs from the quiet Sun sound speed $c_o$ in the following fashion:
\begin{equation}
   \big( c / c_o \big)^2 = 1 + f(\alpha) g(h),
\end{equation}
where $\alpha$ is the angular distance from the center of the active region, $h$ the radial distance from the photosphere, and
\begin{equation}
  f(\alpha) =
  \begin{cases}
    1+\cos( \pi \alpha / \alpha_d ) & \mbox{for~} |\alpha| \le \alpha_d; \\
    0 & \mbox{otherwise}.
  \end{cases}
\end{equation}
The radial profile $g(h)$ of the prescribed sound speed perturbation is shown in Figure~\ref{fg2}.
The profile has been derived by inversions of the time-distance measurements of an actual sunspot~\citep{kos00}, and confirmed by a number of other local helioseismology inversions \citep{Jensen2001,Sun2002,Basu2004,Couvidat2006,Zharkov2007}.
Some of these studies have shown that the significant sound speed perturbation associated with the sunspot structure probably extends deeper than what was originally inferred.
Also, investigations of large active regions by \citet{kosovichev2006} have indicated that the perturbations are extended significantly deeper than those for the relatively small and isolated sunspot in \citet{kos00}.
Therefore, we extended this profile into the deeper layers as shown in Figure~\ref{fg2}.
The simulations have been performed for three different active region horizontal sizes $\alpha_d$ corresponding to radii at the solar surface of 45, 90 and 180~Mm, respectively.
Effects of structure variations with depth or the strength of the perturbations have not been studied.


\section{FAR-SIDE IMAGING PROCEDURE}

\citet{zha07} has imaged the solar far side using medium-$l$ data acquired by SOHO/MDI \citep{scherrer1995}.
MDI medium-$l$ data consist of line-of-sight photospheric velocity images with a cadence of 1 minute and a spatial sampling of $0.6\degr$ per pixel (here and after, degree means heliographic degree).
The data are mapped into heliographic coordinates using Postel's projection, and only the central $120\degr \times 120\degr$ region of the solar disk is used for the far-side imaging analysis.
The observational time series were 2048 minutes long.

From the simulations, very similar datasets were generated.
Radial velocity maps were computed at a location of 300~km above the photosphere, approximately at the formation height of MDI Dopplergrams~\citep{nor06}, and stored with a 1-minute cadence and a spatial resolution of $0.703\degr$ per pixel, slightly lower than the resolution of the MDI data.
The region selected for the analysis was of the same size, $120\degr \times 120\degr$, as in the analysis of the MDI observations.
The first 500 minutes of each simulation were discarded as they represent transient behavior, and the following 1024 minutes were used in the analysis.
This is only half of the duration used in the observational analysis, but as Figure~\ref{fg1} shows, four- and five-skip acoustic signals are sufficiently strong to perform the far-side analysis even with such a relatively short period.

\begin{figure}
  \epsscale{0.9}
  \plotone{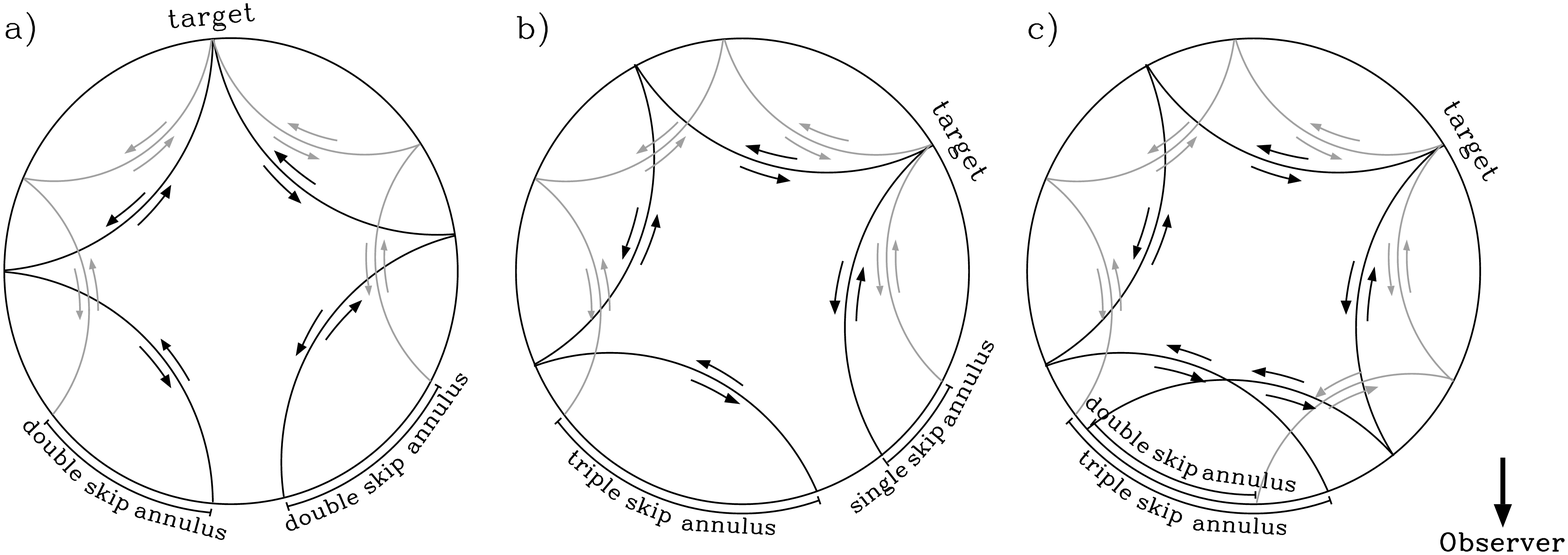}
  \caption{Illustrations of the four-skip and five-skip measurement schemes used
  in the far-side imaging method of~\cite{zha07}. The skips
  correspond to the ray paths of acoustic waves traveling between
  surface points through the solar interior.
Specifically, ({\it a\/}) represents the scheme with two skips on
either side of the target point, ({\it b\/}) the single-triple skip
scheme, and ({\it c\/}) the double-triple skip scheme.}
  \label{fg3}
\end{figure}

The rest of the procedure for the simulation data is the same as for
the observations presented in~\citet{zha07}. After the remapping,
the data are filtered in the Fourier domain, and only waves that
travel long enough to return to the near side from the back side
after four and five rebounces are kept. The time-distance
cross-covariance function is computed for points inside the annuli
as indicated in Figure~\ref{fg3}. The locations
and sizes of these annuli depend on the measurement scheme. For the
four-skip scheme and the double-double skip combination, the annulus
covers a range of distances of $133.8\degr - 170.0\degr$ from the
targeted point on the far side. For the single-triple combination, this range is $66.9\degr - 85.0\degr$ for the single skip, and
$200.7\degr - 255.0\degr$ for triple skip.  For the five-skip
scheme, the annulus covers the range of $111.6\degr - 174.0\degr$
from the targeted point for the double skip, and $167.4\degr -
261.0\degr$ for the triple skip. The four-skip scheme can recover
images of $190\degr$ in longitude ($5\degr$ past the limb to the
solar near side on either limb), while the five-skip scheme recovers
a total of $160\degr$ in longitude, somewhat less than the whole
far side. As usual, the cross-covariance functions for different
distances are combined after appropriate shifts in time based on ray theory predictions.
The final cross-covariance functions are fitted using a Gabor wavelet \citep{kosovichev1997} to derive the acoustic phase travel times for the four- and five-skip schemes separately.
After a mean background travel time is subtracted from each map, the residual travel times maps show variations, corresponding to active regions on the far side.


\section{RESULTS}


\subsection{Sensitivity}

In order to examine the sensitivity of the time-distance far-side
imaging technique to the size of active regions, we have simulated
the global acoustic wave fields for solar models with sound speed
perturbations of 3 different values of their radius: 180~Mm (large),
90~Mm (medium), and 45~Mm (small). The radial structure of the
sound speed perturbation has been given in Sec.~2.

\begin{figure}
  \epsscale{0.9}
  \plotone{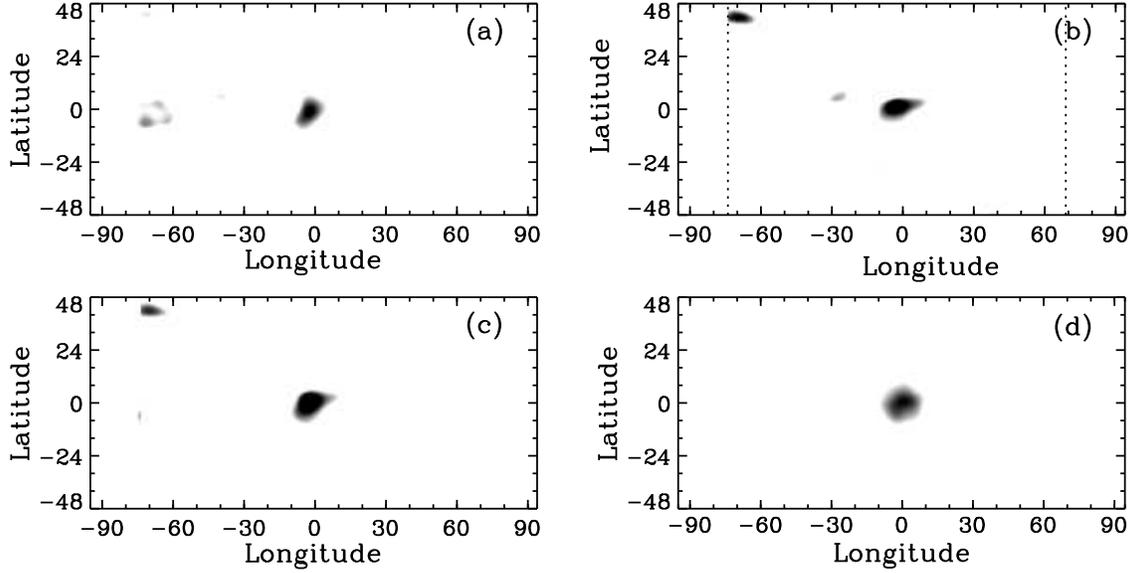}
  \caption{Results for an active region with a radius of 90~Mm positioned
  at the center of the solar far side.
The individual panels show: ({\it a\/}) the far-side image from the
four-skip measurement scheme, ({\it b\/}) the image from the
five-skip scheme, and ({\it c\/}) the combined far-side image. The
images show the derived travel-time signals for the different
schemes, and are displayed with thresholds of $-3.5\sigma$ to
$-2\sigma$, with $\sigma$ being the standard deviation of the
travel time variations (noise level). The dotted lines in panel
({\it b\/}) indicate the spatial limits of the five-skip scheme,
which does not cover the whole far side. As an illustration of the
actual size and location of the model active region, panel ({\it
d\/}) depicts the average acoustic power on the far side computed at
the photospheric level. Inside the active region, a reduction in the
average acoustic power is observed ({\it indicated in black\/})
compared to the quiet Sun ({\it shown in white\/}).}
  \label{fg4}
\end{figure}

\begin{figure}
  \plotone{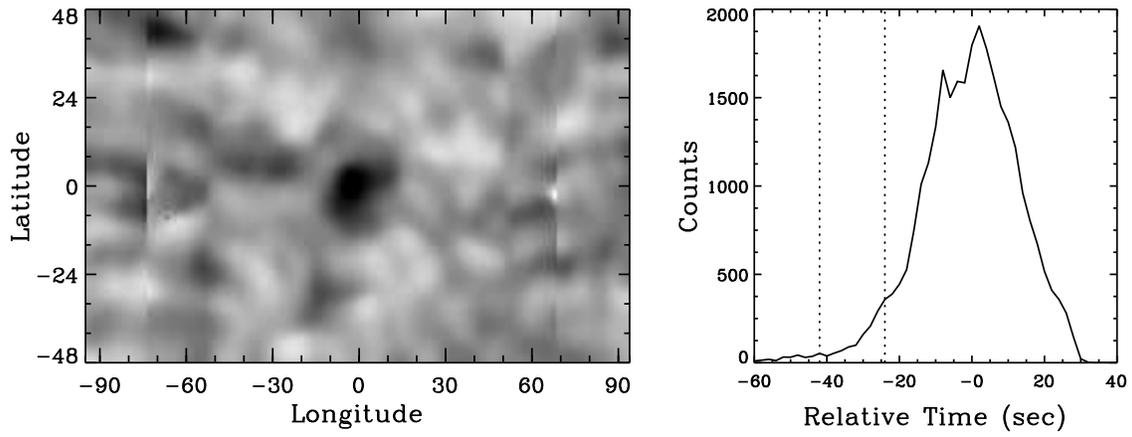}
  \caption{The combined method image from Figure~\ref{fg4}(c) without thresholding ({\it left\/}), and the distribution of travel times residuals in the image ({\it right\/}). The dotted lines indicate the threshold limits used for rendering the image in Figure~\ref{fg4}(c).}
  \label{fg5}
\end{figure}

Figure~\ref{fg4} presents the case when a medium-sized active region
is located at the far-side center (directly opposite to the
observer). It can be seen that both four- and five-skip measurement
schemes can recover this far-side region, but with some level of
spurious features. The combined image from both schemes gives a
better active region image, though not completely clear of spurious
features. The images are displayed with thresholds of $-3.5\sigma$
to $-2\sigma$, where $\sigma$ is the standard deviation of the
travel-time perturbations, 
in order to isolate the strong negative signals associated with
active regions. The original unrestricted image without
thresholding, and the corresponding probability distribution
function of the travel time residuals are shown in Figure~\ref{fg5}.
In this particular case, $\sigma$ is of the order of 12 seconds for the combined image.
For comparison, a lower value of 3.3 seconds was found in observations~\citep{zha07}.
The noise level depends on the stochastic properties of solar waves
and the length of the data time series. This probably explains the
difference in the noise levels. However, this difference is not
significant for this study since we measure the signal relative to
the noise level.

\begin{figure}
  \plotone{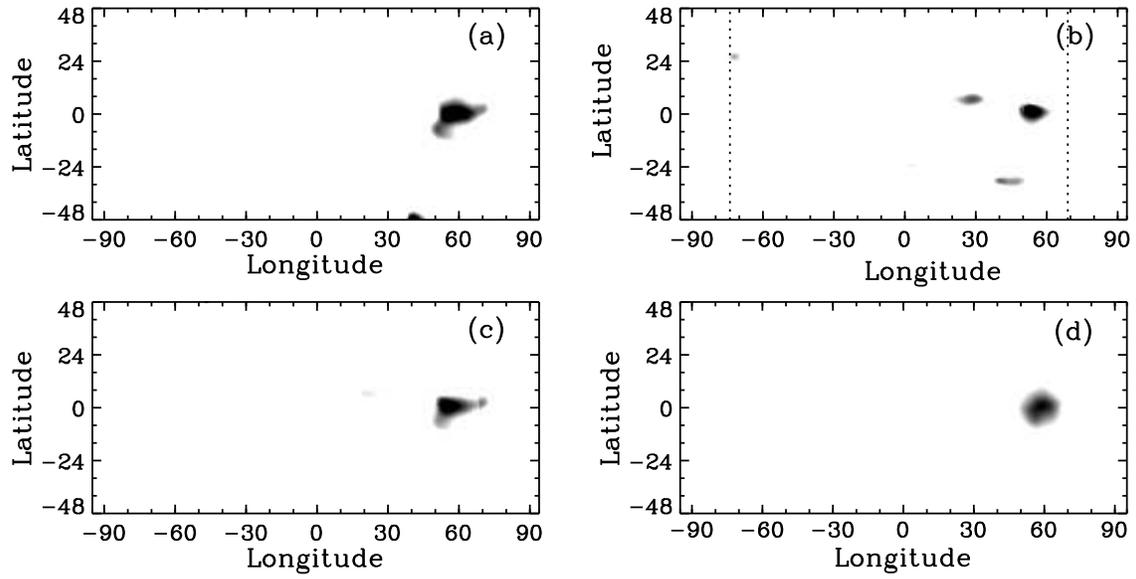}
  \caption{Same as Figure~\ref{fg4}, except the model active region is placed near the limb of the far side.}
  \label{fg6}
\end{figure}

Figure~\ref{fg6} shows the same, medium-sized active region, but now
located closer to the far-side limb. Once again, the combined
far-side image gives the best result. It is clear from both
Figures~\ref{fg4} and \ref{fg6} that the time-distance technique
determines the size and location of the far-side active regions well
but fails to  accurately image their shape.

\begin{figure}
  \plotone{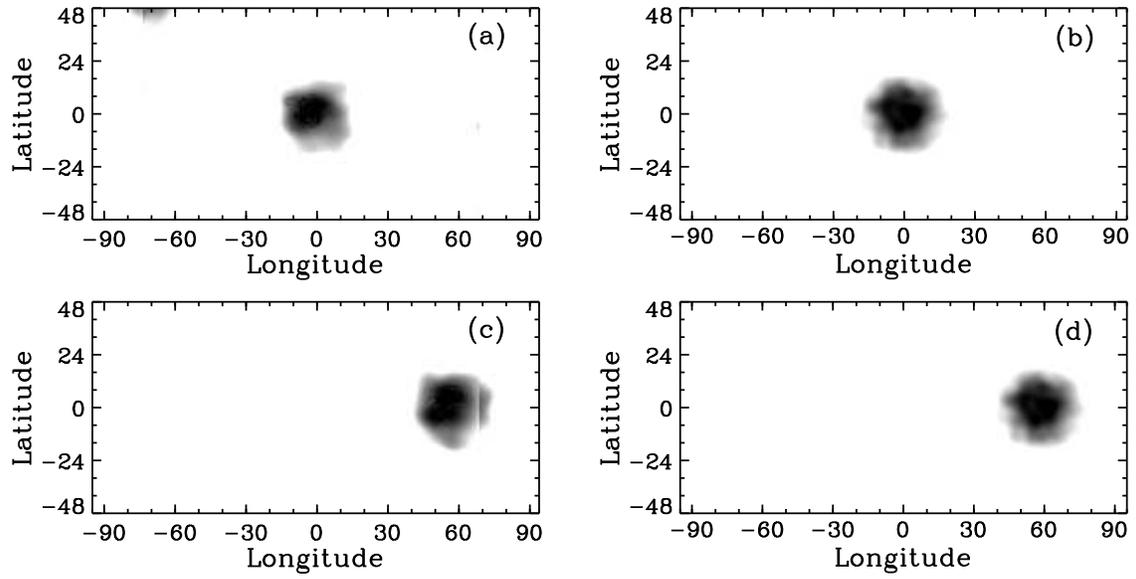}
  \caption{Far-side images of a large model active region with a radius of 180~Mm
  positioned at the center ({\it a\/}) and near the limb ({\it c\/}) of the far side.
The acoustic power at the photosphere for the two cases are shown in
panels ({\it b\/}) and ({\it d\/}), respectively.}
  \label{fg7}
\end{figure}

Figure~\ref{fg7} presents the travel-time images combined from the
four- and five-skip measurements for the simulations of the large
active region. It is evident that the time-distance technique gives
the correct size, location, and even shape of the far-side active
regions for both far-side locations: at the center, and near the
limb.

For the case with a small active region (45~Mm radius), the time-distance helioseismology imaging fails to provide any credible signature of the existence of the region on the far side.
The travel-time maps are not shown for this case, since they don't show any significant features.
Of course, it should come as no surprise that the imaging technique has a lower limit on what size of active region can be detected.
The time-distance far-side imaging method used in this study employs only the oscillation modes with spherical harmonic degrees $l$ between 3 and 50.
It is conceivable that structures comparable in size or even smaller than the horizontal wavelength of the acoustic waves used in the analysis will have little effect on such waves.
Such small structures would be hard or impossible to detect.
A simple estimate of the node-to-node distance for a spherical harmonic of degree 50 (the highest used in the analysis) gives about 90~Mm at the surface, or twice the radius of the small active region.

\begin{figure}
  \plotone{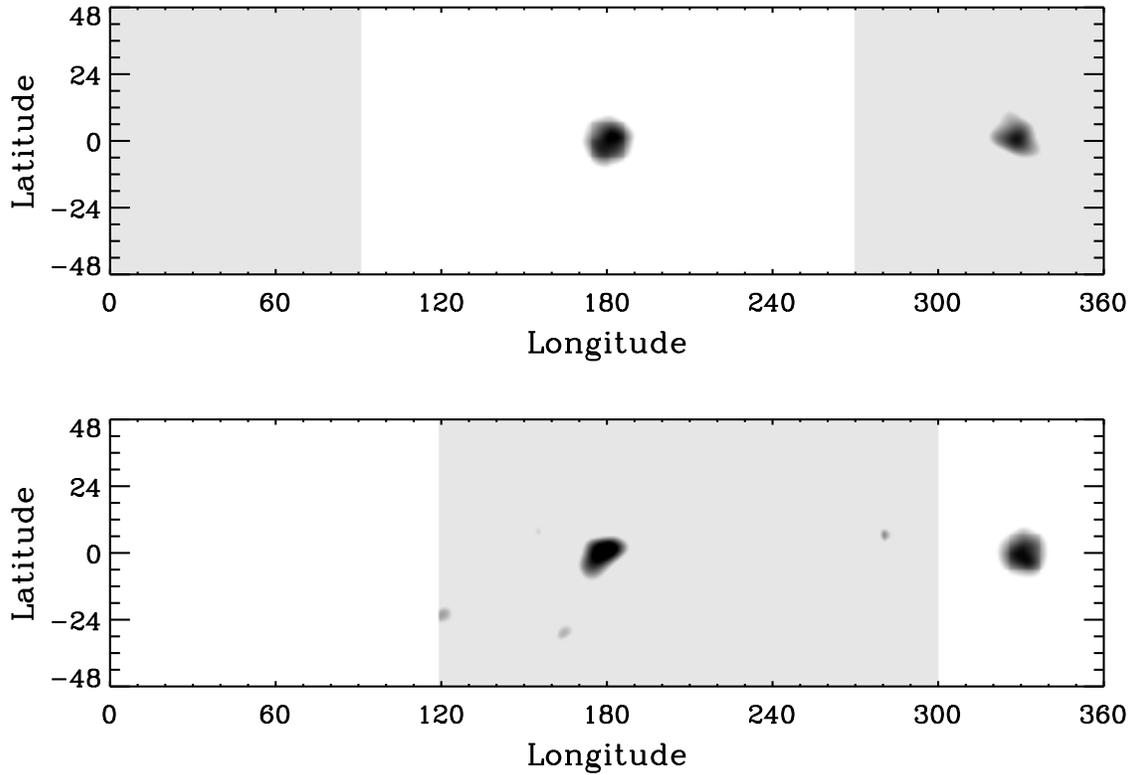}
  \caption{Far-side images for a simulation with two medium-sized model active regions at the equator.
The two regions are $150\degr$ apart longitudinally, namely, at $180\degr$ and $330\degr$.
For the two panels, two different regions have been selected as the near side.
The panels depict a combination of the acoustic power map of the
near-side ({\it white background\/}), and the far-side map ({\it
gray background\/}) computed from the oscillation data on the near
side. }
  \label{fg8}
\end{figure}

It is quite common that multiple active regions are present on the Sun.
Some of them may produce perturbations of the wave field, which may interfere with the perturbation of a targeted active region.
In order to examine whether the different regions would interfere with each other in the far-side images, we performed a simulation with two medium-sized active regions located at the solar equator, $150\degr$ apart from each other.
We have examined various different far-side locations of the active regions, and two examples are presented in Figure~\ref{fg8}.
In all cases we found that these two active regions do not interfere with each other.
Both active regions have been imaged correctly as if they were the sole regions on the Sun, except that some ``ghost images'' appeared under certain circumstances.
However, such artifacts also appear for a single active region case under the same circumstances, as described in the next section.

For convenience, the active regions in the numerical experiments were placed on the far-side equator.
On the real Sun, though, active regions are often far from the equator, and one may be confronted with additional effects such as foreshortening and line-of-sight projection.
However, these effects are expected to be small because only oscillations of relatively low angular degrees are used in the analysis and also
because of the rather small observing window extending only $60\degr$ from the disk center.
To test this expectation, we have performed an additional numerical experiment with a medium-sized active region placed at a latitude of $20\degr$ above the equator, used line-of-sight velocities instead of the pure radial velocities, and included the effect of foreshortening.
The results were not significantly different from those in Figure~\ref{fg4}.


\subsection{Ghost Images}

\begin{figure}
  \plotone{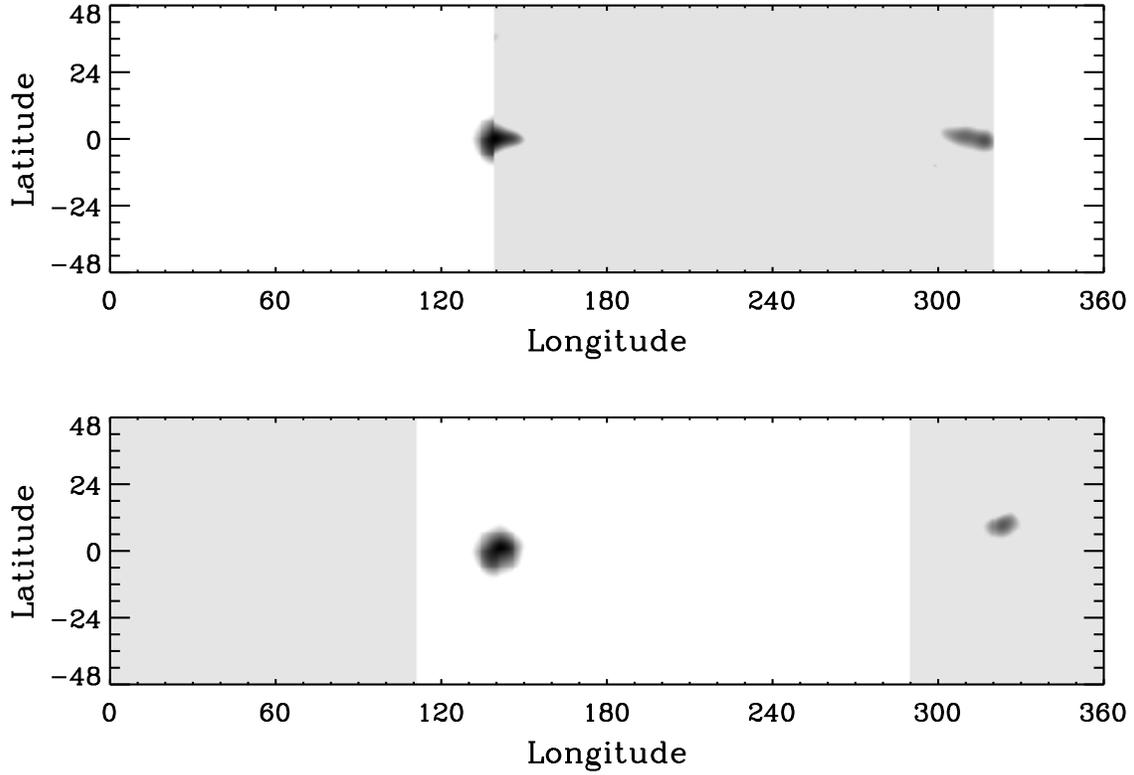}
  \caption{Far-side images for a simulation with a single, 90~Mm-radius model
  active region. The active region is placed at a longitude of $140\degr$.
  Its smaller and weaker ``ghost image'' can be found at approximately $320\degr$
  in both panels. Similarly to Figure~\ref{fg8}, two different parts of the
  surface have been selected as the near side for the two panels, and again
  show a combination of the acoustic power maps of the near side
  ({\it white background\/}) and the far-side image ({\it gray background\/}).}
  \label{fg9}
\end{figure}

It is found that when an active region is placed at certain locations, a ``ghost image'' of the active region may appear in the far-side image.
Figure~\ref{fg9} presents two such examples when an active region on the near side is close to the limb.
A ``ghost image'' appears approximately at the antipode of this active region, with a weaker acoustic travel time signal and smaller in size.
Note that because of the selection of very small $l$'s when computing far-side images, the spatial resolution of images is only about $10\degr$.
Therefore, the ``ghost image'' may appear several degrees away from the antipode of the region.

Given the measurement scheme, it is very reasonable to expect such an artifact when the active region is located close to the near-side limb.
Consider, for example, a single-triple skip combination in the four-skip measurement scheme.
If we select an annulus $70\degr$ from a targeted far-side quiet region, this annulus is also $250\degr$ away from that quiet region's antipode.
If an active region is located there, acoustic waves with travel time deficits caused by that active region are not filtered out because their distance range also falls in the triple-skip range in our analysis (compare annulus radii in Sec.~3).


\section{DISCUSSION}

We have successfully simulated the global acoustic wavefield of the Sun and have used the simulation data to validate the time-distance far-side imaging technique for two measurement schemes with four and five skips of the acoustic ray paths.

We have found that this technique is able to reliably detect our model active regions with radii of 90~Mm and 180~Mm.
The locations and sizes of the far-side active regions are determined correctly, although, their shapes are often slightly different from the original.
Expectedly, larger active regions are easier to detect, and their images are more clear.
For the small active region of 45~Mm radius, the far-side imaging method fails since it is below the resolution limit.
In the case of more than one active regions present on the solar surface, we have found that they do not affect each other's detection.
The time-distance analysis can detect the individual active regions as if they were completely independent.

We have also shown that when an active region is located close to the limb on the near side, a ``ghost image'' may appear in the far-side
image, approximately at its antipode, but relatively weak and smaller in size.
Even though this effect is not completely unexpected, it has not been noticed in previous analyses of observational data in both helioseismic holography~\citep{bra01} and time-distance helioseismology~\citep{zha07}.
This is an important finding and gives us hints on when and where features in observational far-side images may merely be artifacts (i.e. ``ghost images'') and are not caused by actual far-side active regions.


\acknowledgments

We thank Dr. Alan A. Wray and Dr. Konstantin V. Parchevsky for
reading this manuscript and their helpful comments.

This work was supported by NASA's ``Living With a Star'' program.
Support from the NASA Postdoctoral Program administered by Oak Ridge
Associated Universities is gratefully acknowledged.



\clearpage

\clearpage

\clearpage

\clearpage

\clearpage

\clearpage

\clearpage

\clearpage

\end{document}